\documentclass[preprintnumbers, amsmath, amssymb, showpacs, twocolumn, superscriptaddress, nofootinbib]{revtex4-1}

\usepackage{amsmath}
\usepackage{amssymb}
\usepackage{amsfonts}
\usepackage{amssymb}
\usepackage{array}
\usepackage{dcolumn}
\usepackage{bm}
\usepackage{bbm}
\usepackage{cases}
\usepackage{epsfig}
\usepackage{graphicx}
\usepackage{hyperref}
\usepackage{subfigure}
\usepackage{wasysym}
\usepackage{xcolor}
\usepackage{lineno}

\newcommand{\be}{\begin{equation}}
\newcommand{\ee}{\end{equation}}

\newcommand{\bea}{\begin{eqnarray}}
\newcommand{\eea}{\end{eqnarray}}

\def\gsim{ \lower .75ex \hbox{$\sim$} \llap{\raise .27ex \hbox{$>$}} }
\def\lsim{ \lower .75ex \hbox{$\sim$} \llap{\raise .27ex \hbox{$<$}} }

\begin{document}


\title{Searching for the Dark Force with 21-cm Spectrum in Light of EDGES}

\author{Chunlong Li}
\email{chunlong@mail.ustc.edu.cn}
\affiliation{CAS Key Laboratory for Research in Galaxies and Cosmology, Department of Astronomy, University of Science and Technology of China, Hefei 230026, China}
\affiliation{School of Astronomy and Space Science, University of Science and Technology of China, Hefei 230026, China}

\author{Yi-Fu Cai}
\email{yifucai@ustc.edu.cn}
\affiliation{CAS Key Laboratory for Research in Galaxies and Cosmology, Department of Astronomy, University of Science and Technology of China, Hefei 230026, China}
\affiliation{School of Astronomy and Space Science, University of Science and Technology of China, Hefei 230026, China}

\begin{abstract}
The EDGES Collaboration has recently announced the detection of the 21-cm spectrum with an absorption profile centred at $78$ megahertz, of which the depth is deeper than that expected by the standard cosmological paradigm. To enrich the heating process of baryons due to scattering with dark matter during dark ages, we in this Letter explore the possibility of extra heat transfer between dark sector compositions and their observational signatures on the 21-cm cosmological spectrum. By parameterizing interaction models of the dark Universe, we find that the observational constraint on the parameter space of dark matter can be slightly relaxed but the discrepancy with the commonly predicted parameter space of weakly interacting massive particles remains. Our analyses also reveal that the interaction between dark compositions may leave observational signatures on the 21-cm spectrum during dark ages and thus would become detectable in the forthcoming 21-cm cosmology.
\end{abstract}

\pacs{98.80.-k, 98.80.Es, 95.36.+d, 95.36.+x}

\maketitle


\section{Introduction}
Recently, the low-band antenna of the Experiment to Detect the Global Epoch of Reionization Signature (EDGES) reported the detection of an absorption profile in the global 21-cm spectrum centred at $78$ MHz, of which the cosmological redshift corresponds to $z \approx 17$ \cite{Bowman:2018yin}. To compare with the theoretical prediction of the standard cosmological paradigm, EDGES also reported an excess of the signals with the statistic significance at about $3.8\sigma$. It is known that the intensity of the 21-cm line signals relies on the difference between the spin temperature of the hydrogen atoms and that of the Universe, and thus a heat transfer of the hydrogen gas could give rise to an absorption feature in the global 21-cm spectrum due to its cooling process. The probe of these cosmological 21-cm lines from neutral hydrogen are significant to explore the epoch of reionization, which is almost invisible to other astronomical instruments (namely, see \cite{Furlanetto:2006jb, Morales:2009gs, Pritchard:2011xb} for comprehensive reviews).

The studies of the anomaly reported by EDGES have been discussed by introducing extra contribution to the cooling process of the hydrogen gas due to new physics during the early stage of the Universe, namely, the energy transfers between baryon and dark matter (DM) \cite{Tashiro:2014tsa, Barkana:2018lgd, Fialkov:2018xre, Munoz:2018pzp, Berlin:2018sjs, Barkana:2018qrx, Mahdawi:2018euy}, the impacts from star formation \cite{Mirocha:2018cih, Ewall-Wice:2018bzf, Hirano:2018alc} or even intergalactic medium \cite{Venumadhav:2018uwn}, the dynamical dark energy effects \cite{Costa:2018aoy, Hill:2018lfx}, the primordial black holes \cite{Clark:2018ghm, Hektor:2018qqw}, or nonconventional DM scenario \cite{Safarzadeh:2018hhg, Munoz:2018jwq}.
To address the EDGES anomaly of 21-cm intensity, however, there is inevitably the severe fine tuning on the parameter space of the most prevailing DM candidates, which corresponds to the weakly interacting massive particles (WIMPs) in the standard paradigm. Accordingly, the required DM annihilations or decay would lead to an unexpected heating of the hydrogen matter \cite{Fraser:2018acy, DAmico:2018sxd, Cheung:2018vww, Slatyer:2018aqg, Mitridate:2018iag}. Alternatively, it is possible to consider the injection of soft photons to uplift the temperature of background radiation \cite{Fraser:2018acy, Pospelov:2018kdh, Liu:2018uzy}, which implies the possible existence of extra photons \cite{Feng:2018rje} as favored by the ARCADE 2 data \cite{Fixsen:2009xn}.

For models based on the heat transfer between baryonic matter and DM, the DM temperature is much lower than the hydrogen gas since in the standard paradigm DM particles decouple from the thermal bath much earlier than the other. Thus the temperature of hydrogen gas can be reduced via the scattering with DM. However, the parameter space of DM models supported by the EDGES data is severely discrepant with other experimental bounds \cite{Barkana:2018lgd, Berlin:2018sjs, Barkana:2018qrx, Mahdawi:2018euy}.
To alleviate this theoretical difficult, we in this Letter consider the possibility of extra heat transfer between the compositions within the dark sector that can greatly enrich the heating process of baryons through indirect interactions during dark ages. This mechanism can enlarge the parameter space of the regular WIMP paradigm by involving additional interactions. Thus it may provide alternative explanation for the abnormal absorption feature of 21-cm signal announced by EDGES, although the mass and cross section of DM particles are still tightly limited. Furthermore, we examine the impacts of the interactions between dark compositions and find that they could leave potential signatures that are promising to be tested in the future 21-cm experiments.


\section{The model of dark force}
For the explanation to the 21-cm abnormal signal of interactions between baryons and DM, it has been recently pointed out in Ref.~\cite{Munoz:2018pzp} that, in order to be consistent with the stellar-cooling and fifth-force constraints, only part of the DM is allowed to be interacting with baryons. Moreover, the study in Ref.~\cite{Costa:2018aoy} pointed out that the interaction between DM and dark energy could relax the constraint of the Hubble parameter at early times of the Universe and hence may allow the abnormal absorption signal of 21-cm lines at cosmic dawn. Then, these interesting observations naturally lead to a question as follows. Is it possible for the rest part of the dark sector in the Universe to exert extra effects on the process of heat transfer for baryons? This possibility could occur if the heat transfers between different DM compositions or even between DM and dark energy are permitted. If so, the temperature evolutions of the dark sector needs to be revisited, and because of the interaction between DM and baryons, the same thing would also happen to the hydrogen gas, which shall leave additional influence on the 21-cm spectrum. Consequently, it is necessary to analyze the possible effects of heat transfer caused by the dark forces, i.e., the interactions between different compositions in the dark sector of the Universe.

To depict the interactions within the dark sector, 
we use the subscripts $\chi$ to label the part of dark matter that is allowed to be interacting with baryons and $\psi$ to label the another specy of dark components. We adopt $\rho_{\psi k}$, $p_{\psi k}$ to label the energy density and pressure corresponding to the thermal motion of the hidden dark composition ($\rho_{\chi k}$ and $p_{\chi k}$ have the same meanings for $\chi$) and for their specific meanings we refer readers to the appendix section.
Different from the regular baryon-DM interactive model, the newly introduced $\psi$ field can only interact with the $\chi$ DM that weakly couples to baryons. Thus, the energy conservation equations of the dark sector can be written as
\begin{align}
 \dot{\rho}_{\chi k}+3H(\rho_{\chi k}+p_{\chi k}) &= -Q+n_{\chi} \dot{Q}_{\chi} ~, \nonumber \\
 \dot{\rho}_{\psi k}+3H(\rho_{\psi k}+p_{\psi k}) &= Q ~,
\label{m2}
\end{align}
where $n_{\chi}$ is the number density of $\chi$ particles and $\dot{Q}_\chi$ depicts the rate of heat transfer received by DM due to the interaction between $\chi$ and baryons. The dark force existing in the dark sector is realized by the $Q$ term, where we parametrize its form to be
\begin{align}
\label{modelQ}
 Q=\xi H(\rho_{\chi k}-\rho_{\psi k}) ~.
\end{align}
In this model we have introduced a dimensionless parameter $\xi$ to characterize the dark force, of which the value is expected to determined observationally, and $H$ is the Hubble parameter that measures the expanding rate of the Universe \footnote{Models of DM that couple to dark energy were extensively studied in the literature, namely see \cite{Farrar:2003uw} for the pioneer study and see \cite{Wang:2016lxa} for a comprehensive review and references therein for related analyses.}. Eq. \eqref{m2} can be casted into the standard form of the first law of thermodynamics \cite{Ma:1995ey} with the temperature evolution equations of the dark sector to be improved as below,
\begin{align}
 (1+z)\frac{dT_\chi}{dz} &= (2+\xi)T_\chi-\xi T_{\psi}-\frac{2\dot{Q}_\chi}{3H} ~, \nonumber \\
 (1+z)\frac{dT_{\psi}}{dz} &= (2+\xi)T_{\psi}-\xi T_\chi ~.
\label{s4}
\end{align}
Afterwards, one may choose appropriate values of $\xi$ and the regular interaction model $\bar{\sigma}=\sigma_0v^{-4}$ (For the constraints on this kind of models, we refer readers to the study \cite{Dvorkin:2013cea}.), and then solve Eq.~\eqref{s4} by combining the evolution equations in the traditional baryon-DM interactive model starting from the recombination moment with the baryons tightly coupled to the photon fluid ($T_b=T_{\gamma}$) and with the perfect cold dark sector ($T_{\chi}=0$ and $T_{\psi}=0$, which implies that $\xi$ should be positive to make sure the $\psi$'s temperature to be positively definite) to obtain the evolution of the gas temperature \cite{Munoz:2015bca}.

\section{Analyses and results}
The brightness temperature of 21-cm line $T_{21}$ is expressed as the difference between the spin temperature $T_S$ and the background radiation temperature $T_\gamma$ \cite{Ciardi:2003hg, Zaldarriaga:2003du}, i.e.
\begin{align}
 T_{21}=\frac{3}{32\pi}\frac{T_*}{T_s}n_{HI}\lambda_{21}^3\frac{A_{10}}{H}\frac{T_S-T_\gamma}{1+z} ~.
\label{T21}
\end{align}
where $T_* = 5.9 \mu eV$ is the energy corresponding to the 21-cm transition, $A_{10}$ is the downward spontaneous Einstein coefficient of the 21-cm transition \cite{AliHaimoud:2010ab, AliHaimoud:2010dx}, $n_{HI}$ is the number density of neutral hydrogen, $\lambda_{21}$ is the wavelength of 21-cm line. For the dark age the expression of spin temperature $T_S$ is
\begin{align}
 T_S=T_\gamma+\frac{C_{10}(T_b-T_\gamma)}{C_{10}+A_{10}\frac{T_b}{T_*}} ~,
\end{align}
where $C_{10}$ is the collisional transition rate \cite{Lewis:2007kz, Furlanetto:2007te}.

\begin{figure}
{\includegraphics[width=.45\textwidth]{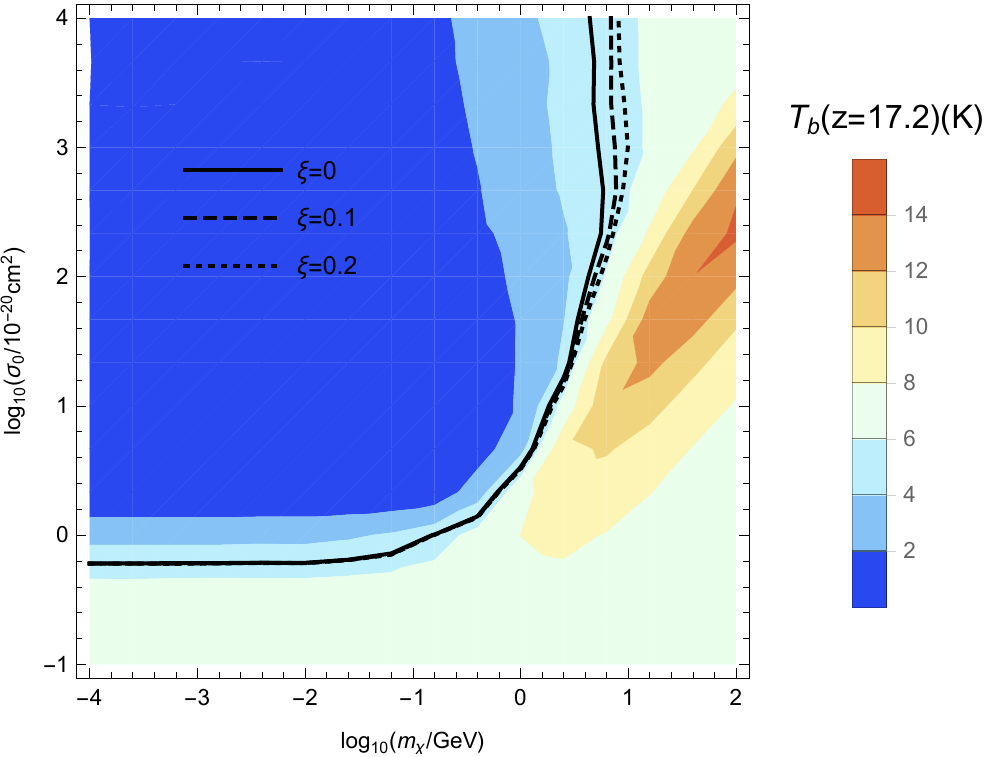}}
\caption{Constrains on the DM parameter space using the gas temperature at $z=17.2$. In the contour plot of the gas temperature $T_b$, the isothermal curves of $T_b=5.1~{\rm K}$ with different values of the dark force parameter $\xi$ are presented.}
\label{fig:a}
\end{figure}

The observed absorption feature at cosmic dawn originates from the indirect coupling of $T_S$ to the gas temperature $T_b$ by stellar Lyman-$\alpha$ photons \cite{Chen:2003gc} via the Wouthuysen-Field effect \cite{Wouthuysen:1952, Field:1958}, resulting in $T_S \approx T_b$. This is why the 21-cm line abnormal absorption signal, which is a factor of two deeper than expected at $z=17$ \cite{Bowman:2018yin}, would arise naturally if the baryons had a lower temperature than the case in the standard cosmological paradigm. Consequently, we in Fig. \ref{fig:a} use the gas temperature to reflect the absorption intensity of the 21-cm line at cosmic dawn with different value choices of the parameter $\xi$.

At the redshift $z=17.2$, the observed minimum absorption at a $99\%$ confidence level implies the temperature of the gas is less than $5.1~{\rm K}$. In Fig. \ref{fig:a} we numerically show the different isothermal curves at $z=17.2$ in the cross-section versus DM particle mass coordinate with and without introducing the dark interaction. The region that is on the right and lower part of the $T_b=5.1~{\rm K}$ line without $\xi$ is almost excluded in \cite{Barkana:2018lgd}. However, if the parameter $\xi$ is not vanishing, one can find that the constraint on the DM particle mass is relaxed while the constraint on the cross section is barely altered.

\begin{figure}
{\includegraphics[width=.45\textwidth]{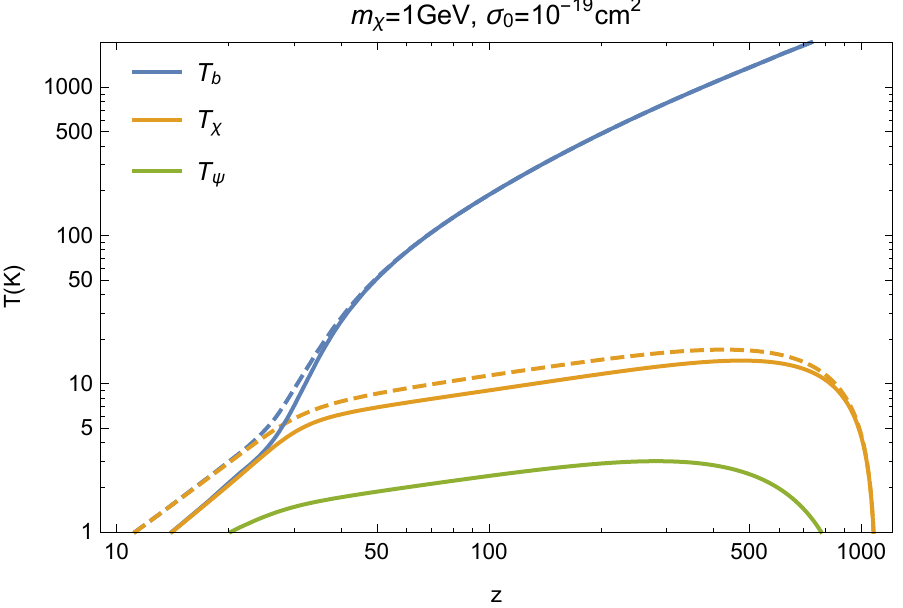}}
{\includegraphics[width=.45\textwidth]{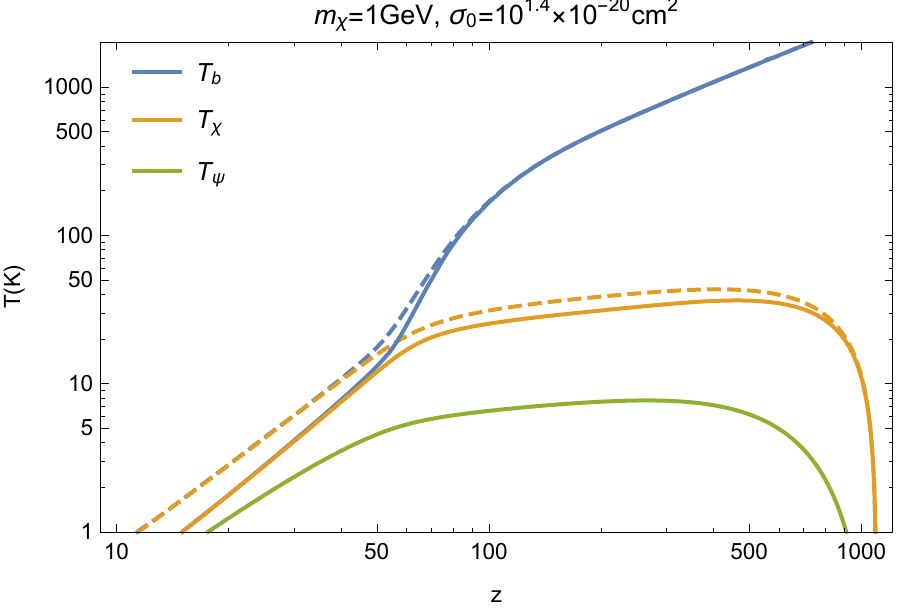}}
{\includegraphics[width=.45\textwidth]{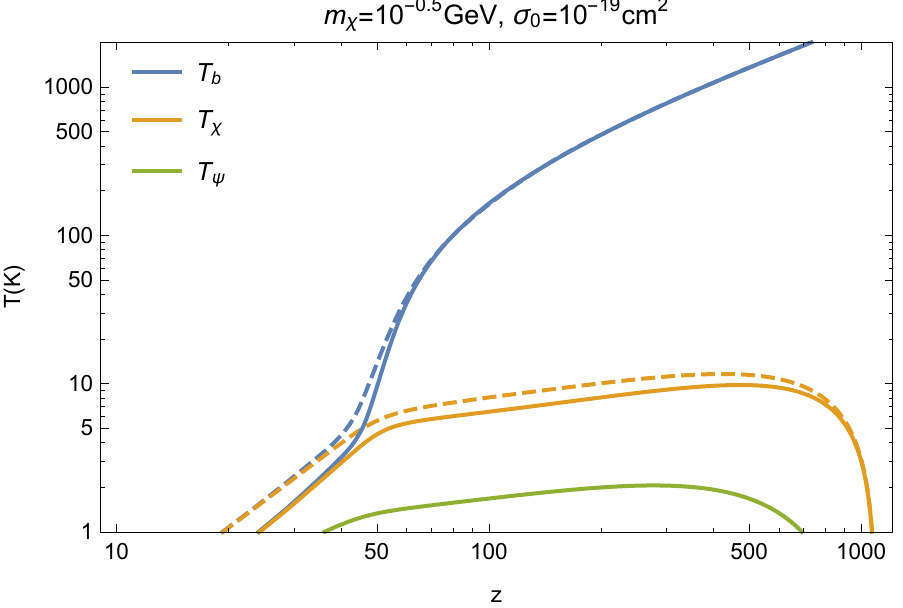}}
\caption{The temperature evolutions for different matter components as functions of the redshift $z$. The dashed line represents the traditional baryon-DM interactive model and the solid line corresponds to our model with the dark force parameter $\xi=0.6$.}
\label{fig:b}
\end{figure}

In our model, due to the fact that DM can transfer heat to another dark sector composition, the temperature of DM may be suppressed, which means in the same condition we will get a lower gas temperature. This is why a non-zero $\xi$ could change the parameter space that has been ruled out. For illustration, we in Fig. \ref{fig:b} numerically plot the evolutions of temperature of the hydrogen gas, the $\chi$ DM field and the underlying $\psi$ dark sector for different values of the mass as well as cross section of the $\chi$ DM particles. One can easily find that, due to the presence of the additional heat transfer, the temperature of gas and DM is suppressed. Moreover, the deviation of the gas temperature from the standard evolution mainly focuses on the low redshift, which means the heat transfer with the additional dark sector can exert the obvious effects on the 21-cm signal of cosmic dawn instead of dark ages. 

\begin{figure}
{\includegraphics[width=.45\textwidth]{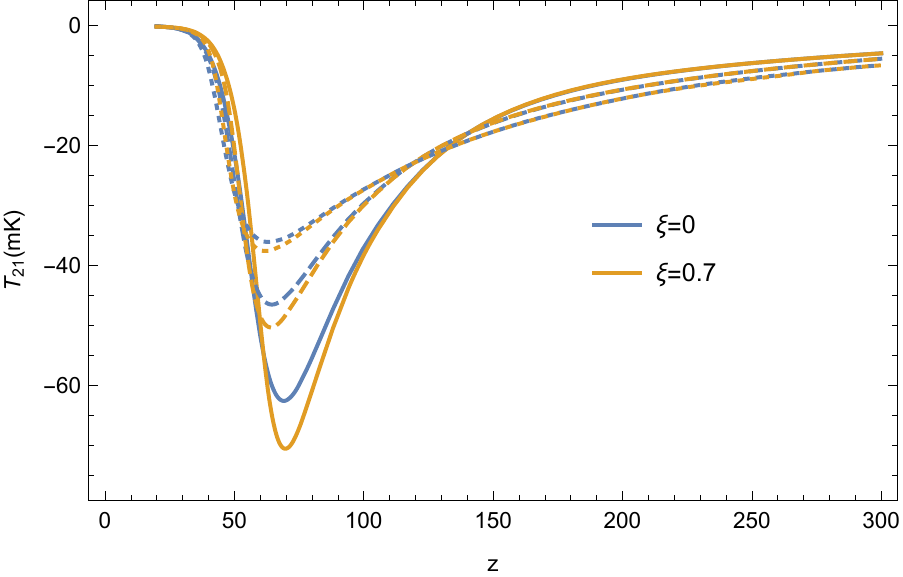}}
\caption{Evolutions of the brightness temperature (\ref{T21}) of the 21-cm line during dark ages. The DM mass $m_{\chi}$ is fixed as $1 GeV$. The dotted, dashed and solid curves correspond to the parameter choices of $\sigma_0=10^{1.2} \times 10^{-20} {\rm cm}^2$, $10^{1.3} \times 10^{-20} {\rm cm}^2$ and $10^{1.4} \times 10^{-20} {\rm cm}^2$, respectively.}
\label{fig:c}
\end{figure}

However, as shown in the second and last panel of Fig. \ref{fig:b}, if we enhance the interaction between DM and baryons, i.e. decrease the mass of DM particle or increase the cross section, the deviation of the gas temperature from the standard evolution would have occurred at earlier moments, and thus could also have impacts on the 21-cm signal of dark ages. 
In order to show this point clearly, we numerically plot the 21-cm signal at dark ages with different value choices of the cross section and the parameter $\xi$ in Fig. \ref{fig:c}. For each cross section we have considered two value choices of $\xi$, and then we can clearly see that the influence of the dark force can only becomes manifest when the signal is relatively large, which is consistent with the previous statements. 

\begin{figure}
{\includegraphics[width=.45\textwidth]{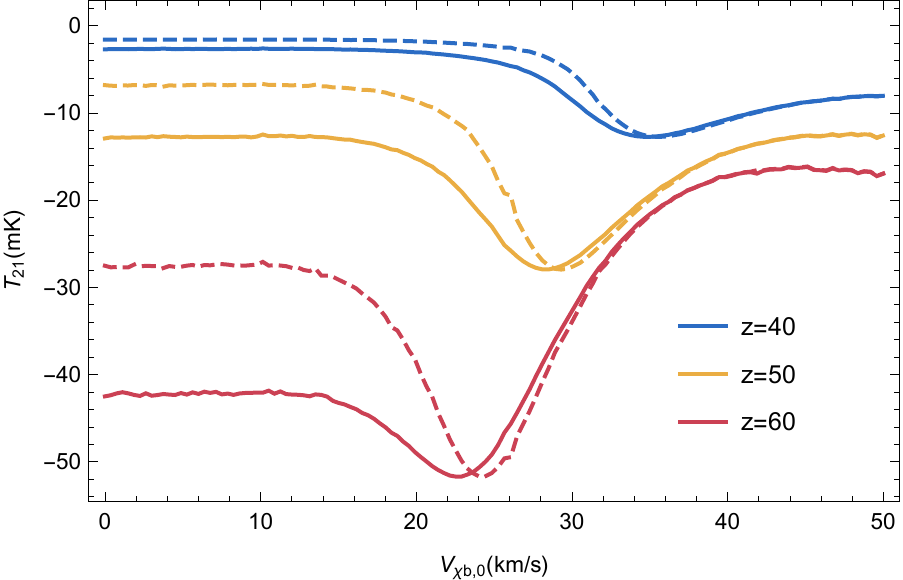}}
\caption{Brightness temperarure $T_{21}$ of the 21-cm line for $m_{\chi}=1GeV$ and $\sigma_0=10^{1.2}\times 10^{-20}cm^2$ for $\xi=0$ in solid curve and $\xi=0.7$ in dashed curve at redshift $z=40$, $z=50$ and $z=60$, respectively.}
\label{fig:d}
\end{figure}

In the above discussion, we take initial relative velocity $V_{\chi b,0}$ between DM and baryons as the root-mean-square velocity $29~{\rm km/s}$ \cite{Tseliakhovich:2010bj, Ali-Haimoud:2013hpa}. However, given that the fluctuation of 21-cm line brightness temperature is determined by the fluctuation of initial relative velocities \cite{Ali-Haimoud:2013hpa}, it's necessary to study the change of the dependence on initial relative velocities for 21-cm line brightness temperature in the presence of the extra heat transfer parameter $\xi$. In Fig. \ref{fig:d}, we show $T_{21}$ as a function of $V_{\chi b,0}$ with different values of $\xi$ and redshift. We can find that in small velocity regions the brightness temperature will be weakened by the presence of $\xi$, which will smooth the fluctuations of the brightness temperature for 21-cm line. So we expect the presence of the underlying heat transfer in the dark sector of the Universe will leave an observable effect on power spectrum of 21-cm fluctuations.

\section{Conclusions}
In this Letter we study the 21-cm line signals due to the possible interaction of heat transfer between dark sector compositions. One composition is the part of DM that can interact with the ordinary matter through the interaction cross section $\bar{\sigma}=\sigma_0v^{-4}$, the other composition is some invisible part of the dark sector which can only exchange heat with the former DM. This part of dark sector can be either dark energy or some unknown species of DM, and hence we start from a rather general model and derive the modified evolution equation of temperature for different compositions. The results show that the presence of this underlying heat transfer can relax the constraint on the particle mass of DM for the model $\bar{\sigma}=\sigma_0 v^{-4}$. We further study the change of the temperature evolution for different compositions and find that the deflect of gas temperature evolution from the standard situation mainly focuses on low redshift. However if the interaction between DM and baryons is strong enough, this kind of deflect could also extend to the dark age. So we displayed the 21-cm absorption signal of the dark age for different interaction strength in the presence of the underlying heat transfer. The result shows that the stronger the absorption signal is the greater the influence of the additional heat transfer would be. Finally, by virtue of the change of the 21-cm brightness temperature as the initial relative velocity between DM and baryons, we find that the presence of the additional heat transfer process in dark sector will smooth the fluctuations of 21-cm intensity, which may leave an observable effect on the power spectrum of fluctuations.

We end by discussing the possible observational constraints from other cosmological experiments on the model under consideration. If one treat $\psi$ as dark energy, our model can reduce to the interacting dark energy models and $\psi$ is often regarded as field condensate, which means the thermal motion is typically negligible. Accordingly, this would impose a very severe constraint on the interacting dark energy model from the present cosmological observations. However, it is interesting to note that $\psi$ may be regarded as a second component of dark matter, namely, a model of multiple dark matter species, which was motivated by the theoretical construction of dynamical balance between various dark matter species as studied in \cite{Dienes:2012yz} and by the astronomical constraints from the structure formation as analzyed in \cite{Baldi:2012ua, Baldi:2012kt}. In particular, due to the interaction term, there exists an appearance of isocurvature modes at late times that can lead to nontrivial astronomical features, namely, a fragmentation of collapsed halos, and thus is of observational interest. For our model, this can impose a bound on the $\xi$ parameter  which requires $\xi$ to be no more than unity roughly. Thus, the parameter space considered in the present study is compatible with this astronomical limit. However, it remain necessary to present a more detailed study on the parameter space of our model by combining all relevant astronomical bounds, which is the goal of the following up work. 


\section*{Acknowledgments}
We are grateful to Andrea Addazi, Yong Cai and Antonino Marciano for valuable discussions.
This work is supported in part by the National Youth Thousand Talents Program of China, by the NSFC (Nos. 11722327, 11653002), by the CAST Young Elite Scientists Sponsorship (2016QNRC001), and by the Fundamental Research Funds for the Central Universities.
All numerical computations were operated on the computer cluster LINDA in the particle cosmology group at USTC.


\section*{Appendix}

The appendix section is devoted to the derivation of the evolution equations for the temperatures of the dark sector. We begin with a regular cosmological model that mainly involves baryons and the $\chi$ DM. Their energy densities and pressures are denoted by $\rho_b$, $p_b$ and $\rho_\chi$, $p_\chi$, respectively. In this standard paradigm, one can straightforwardly write down the energy conservation equations as follows,
\begin{align}
 \dot\rho_b +3H(\rho_b+p_b) &= n_b\dot{Q}_b+n_b\dot{Q}_c ~, \nonumber \\
 \dot\rho_\chi +3H(\rho_\chi+p_\chi) &= n_{\chi}\dot{Q}_\chi ~,
\end{align}
where $\dot{Q}_b$ denotes the rate of heat transfer from the baryons and $\dot{Q}_\chi$ represents the rate of heat transfer received by the $\chi$ DM particles. The coefficient $\dot{Q}_c$ depicts the energy transfer due to the Compton scattering with photons, $n_i$ is the number density of each component which evolves as $n_i \propto a^{-3}$.

In usual, the DM and baryons are treated to be ``cold'', which means that their thermodynamic random motions are vanishing and hence are pressureless. However, in order to precisely describe the temperature evolutions of all matter components during dark ages, at the moment the pressures cannot be ignored. According to relativity, the particle energy can be decomposed as $E = mc^2 +m\bar{v^2}/2 +...$, where $m\bar{v^2}/2$ represents the average kinetic energy of thermal motion, which allows us to do the decomposition to be:
$\rho_b = \rho_{b0} +\rho_{bk}$,
$\rho_\chi = \rho_{\chi0} +\rho_{\chi k}$,
$p_b = p_{bk}$,
$p_\chi = p_{\chi k}$,
where $\rho_{b0}, \rho_{\chi0}$ is the background energy density and $\rho_{bk}$, $\rho_{\chi k}$, $p_{bk}$, $p_{\chi k}$ corresponds to the energy density and pressure of thermal motion.

Given that the interaction between DM and baryons and Compton scattering only transfer energy at the level of thermal motion, one can further decompose the energy conservation equations to be the background and the thermal parts, which are given by,
\begin{align}
 \dot\rho_{b0}+3H\rho_{b0} = 0 ~,~~
 \dot\rho_{\chi0}+3H\rho_{\chi0} = 0 ~,
\end{align}
and
\begin{align}
 \dot\rho_{bk}+3H(\rho_{bk}+p_{bk}) &= n_b\dot{Q}_b+n_b\dot{Q}_c ~, \nonumber \\
 \label{b2}
 \dot\rho_{\chi k}+3H(\rho_{\chi k}+p_{\chi k}) &= n_{\chi}\dot{Q}_\chi ~,
\end{align}
respectively.
Since $\rho_{bk}$, $\rho_{\chi k}$, $p_{bk}$, $p_{\chi k}$ describe the thermal motion of the matter particle, by applying the property of ideal gas one can get
\begin{eqnarray}
 & \rho_{bk} = \frac{3}{2}n_bkT_b ~,~ p_{bk} = n_bkT_b ~, \nonumber\\
 & \rho_{\chi k} =\frac{3}{2}n_{\chi}kT_{\chi} ~,~ p_{\chi k} = n_{\chi}kT_{\chi} ~.
\label{i2}
\end{eqnarray}

We adopt the natural units where the Boltzmann constant $k=1$ and speed of light $c=1$. Then from \eqref{b2}, one can derive the temperature evolution equations to be
\begin{align}
\label{s1}
 (1+z)\frac{dT_b}{dz} &= 2T_b-\frac{2\dot{Q}_b}{3H}-\frac{2\dot{Q}_c}{3H} ~, \\
\label{ss}
 (1+z)\frac{dT_\chi}{dz} &= 2T_\chi-\frac{2\dot{Q}_\chi}{3H} ~, 
\end{align}
with the Compton scattering rate $\dot{Q}_c$ being \cite{AliHaimoud:2010dx}
\begin{align}
 \dot{Q}_c=-\frac{4\sigma_T a_r T_\gamma^4 x_e(T_b-T_{\gamma})}{(1+f_{He}+x_e)m_e} ~,
\end{align}
where $T_\gamma$ is the temperature of CMB photons which evolves as $T_\gamma=2.725(1+z)$, $\sigma_T$ is the Thomson scattering cross section, $a_r$ is the Stefan constant, $f_{He}$ is the He-H ratio by number of nuclei, $x_e$ is the free-electron fraction $x_e=n_e/n_H$, it evolves as \cite{AliHaimoud:2010dx}
\begin{align}
 \frac{dx_e}{da}=-\frac{C_P}{aH} \big[ n_HA_Bx_e^2-4(1-x_e)B_Be^{3E0/4T_\gamma} \big] ~,
\label{s2}
\end{align}
where $E_0$ is the ground energy of hydrogen, $C_P$ is the Peebles factor, $A_B$ and $B_B$ are the effective recombination coefficients and the effective photoionization rate to and from the excited state, respectively. For their expression and values we refer readers to \cite{AliHaimoud:2010ab, AliHaimoud:2010dx}.

For the expressions of $\dot{Q}_b$ and $\dot{Q}_\chi$, we choose the model that recently be used to constrain the parameters of DM \cite{Barkana:2018lgd}, where the interaction cross section is parametrized as $\bar{\sigma}=\sigma_0v^{-4}$ and $v$ is the relative velocity of the two particles. We refer to \cite{Munoz:2015bca} for more details, and here we only list the expressions relevant to our calculations. The expression of $\dot{Q}_b$ is given by
\begin{align}
 \dot{Q}_b = \frac{2m_b\rho_\chi\sigma_0e^{-\frac{r^2}{2}}(T_{\chi}-T_b)}{(m_\chi+m_b)^2\sqrt{2\pi}u^3_{th}} 
 +\frac{\rho_\chi}{\rho_m}\frac{m_\chi m_b}{m_{\chi} +m_b}V_{\chi b}D(V_{\chi b}) ~, \nonumber
\end{align}
where $m_b$, $m_\chi$ are the masses of baryons and DM particle, respectively. $\rho_\chi$, $\rho_m$ are their densities, $r \equiv V_{\chi b}/u_{th}$, $u_{th}^2\equiv T_b/m_b+T_\chi/m_\chi$. $V_{\chi b}$ is the relative velocities between DM and baryons. In addition, $D(V_{\chi b})$ is the drag term, of which the form is expressed as
\begin{align}
 D(V_{\chi b}) &= \frac{\rho_m\sigma_0}{m_b+m_{\chi}}\frac{1}{V^2_{\chi b}}F(r) ~, \nonumber\\
 F(r) &\equiv {\rm erf}({r}/{\sqrt{2}}) -\sqrt{{2}/{\pi}} r e^{-r^2/2} ~.
\end{align}
We note that $\dot{Q}_\chi$ can be obtained from the following conservation equation:
\begin{align}
 n_{\chi} \dot{Q}_{\chi} + n_b \dot{Q}_b - \frac{\rho_\chi\rho_b}{\rho_m}D(V_{\chi b})V_{\chi b}=0 ~.
\end{align}

In this work we characterize the interaction inside the dark sector by modifying the energy conservation equations as follows (also see \eqref{m2}),
\begin{align}
 \dot{\rho}_{\chi k}+3H(\rho_{\chi k}+p_{\chi k}) &= -\xi H(\rho_{\chi k}-\rho_{\psi k}) ~, \nonumber \\
 \dot{\rho}_{\psi k}+3H(\rho_{\psi k}+p_{\psi k}) &= \xi H(\rho_{\chi k}-\rho_{\psi k}) ~.
\end{align}
Then, to combine the above equations with \eqref{b2} and to apply the similar expressions for $\psi$ as \eqref{i2}, the evolution equation for the temperature of the $\chi$ DM can be improved from \eqref{ss} to be in form of:
\begin{align}
 (1+z)\frac{dT_\chi}{dz} = (2+\xi)T_\chi -\xi f T_{\psi}-\frac{2\dot{Q}_\chi}{3H} ~, \nonumber
\end{align}
where $f=n_{\psi}/n_{\chi}$ represents the ratio of number densities of these two dark components. 
Additionally, the temperature evolution equation of the extra dark component $\psi$ that only couples to the $\chi$ field is then written as
\begin{align}
 (1+z)\frac{dT_{\psi}}{dz} = (2+\xi)T_{\psi} -\xi f^{-1} T_\chi ~. \nonumber
\end{align}
In this Letter we only consider the simplest case $f=1$.
As a result, one obtains the evolution equations for the temperatures of the dark sector as shown in \eqref{s4}.


\end{document}